\documentclass{elsart}
\usepackage{epsfig}

\def\cal{\mathcal}

\def\to{\rightarrow}

\begin{document}
%\doubles
\begin{frontmatter}

\begin{flushleft}\includegraphics[width=3.5cm]{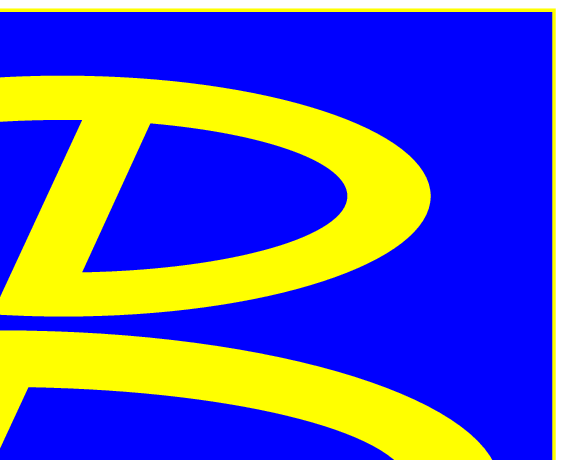}\end{flushleft}
\vspace{-2.5cm}
\vbox{\normalsize%
\noindent%
\rightline{\hfill {KEK preprint 2002-31}}\\
\ \\
\rightline{\hfill {Belle preprint 2002-15}}
}

\vspace{2.0cm}

\title{\boldmath
Measurement of $\chi_{c2}$ Production in Two-Photon Collisions}

\vspace{-0.5cm}

\author{Belle Collaboration}

%\begin{center}
%{\normalsize
%  S.~Uehara           % KEK
% }\end{center}

%\small
%\begin{center}
%\address{
%{High Energy Accelerator Research Organization (KEK), Tsukuba}\\
%}

% This is the e+e- -> e+e- chi_c2 paper's author list.
% Non-responding authors or those who said NO are commented out.
% ====================================================================
% Click the RELOAD button on your web browser to see the updated file.
% ====================================================================
\normalsize
\begin{center}
  K.~Abe$^{8}$,               % KEK
  K.~Abe$^{40}$,              % TohokuGakuin
% N.~Abe$^{43}$,              % TIT
  R.~Abe$^{28}$,              % Niigata
  T.~Abe$^{41}$,              % Tohoku
  I.~Adachi$^{8}$,            % KEK
  Byoung~Sup~Ahn$^{15}$,      % Korea
  H.~Aihara$^{42}$,           % Tokyo
  M.~Akatsu$^{21}$,           % Nagoya
% M.~Asai$^{9}$,              % Hiroshima
  Y.~Asano$^{47}$,            % Tsukuba
  T.~Aso$^{46}$
% ...
\end{center}

\small
\begin{center}
\address{
% ...
$^{8}${High Energy Accelerator Research Organization (KEK), Tsukuba}\\
% ...
$^{15}${Korea University, Seoul}\\
% ...
$^{21}${Nagoya University, Nagoya}\\
% ...
$^{28}${Niigata University, Niigata}\\
% ...
$^{40}${Tohoku Gakuin University, Tagajo}\\
$^{41}${Tohoku University, Sendai}\\
$^{42}${University of Tokyo, Tokyo}\\
% ...
$^{46}${Toyama National College of Maritime Technology, Toyama}\\
$^{47}${University of Tsukuba, Tsukuba}\\
% ...
}
\end{center}

\normalsize

\normalsize
\begin{abstract}
The production of the $\chi_{c2}$  charmonium state in 
two-photon collisions has been measured with the Belle detector
at the KEKB $e^+e^-$ collider. A clear signal for 
$\chi_{c2} \to \gamma J/\psi$, $J/\psi \rightarrow \ell^{+}\ell^{-}$
is observed in a 32.6~fb$^{-1}$ data sample accumulated at
center-of-mass energies near 10.6~GeV, and  
the product of its two-photon decay width and branching fraction
is determined to be
$\Gamma_{\gamma\gamma}(\chi_{c2}){\cal B}(\chi_{c2} \to \gamma J/\psi)
{\cal B}(J/\psi \to \ell^+\ell^-)= 13.5 \pm 1.3(stat.) \pm 1.1(syst.)$~eV.
\end{abstract}

\begin{keyword}
two-photon collisions, \ charmonium, \ $\chi_{c2}$, \ partial decay width \
\PACS 13.20.Gd \   13.40.Hq \   13.65.+i \   14.40.Gx
\end{keyword}

\end{frontmatter}

\clearpage

\section{Introduction}
   The two-photon decay widths ($\Gamma_{\gamma\gamma}$)
of the even charge-parity 
charmonium states provide valuable information for testing
models describing the nature of heavy quarkonia.
Various theoretical calculations describing 
the quark-antiquark system
predict the value of $\Gamma_{\gamma\gamma}(\chi_{c2})$ 
to be within the
range 0.2 -- 0.8 keV \cite{LEP}.
A precise experimental determination of 
$\Gamma_{\gamma\gamma}(\chi_{c2})$
will provide a strong constraint on these models.
The ratio of the two-gluon decay width to
the two-photon decay width
$\Gamma_{gg}(\chi_{c2})/\Gamma_{\gamma\gamma}(\chi_{c2})$ has been
calculated within the framework of perturbative QCD with
first-order correction \cite{thrat} and the result gives
$\Gamma_{\gamma\gamma}(\chi_{c2})=0.47 \pm0.13$~keV.\footnote{This
value is derived from
measurements \cite{pdb} using the assumption that
$\Gamma_{gg}(\chi_{c2})$ is given by $\Gamma(\chi_{c2}
\to {\rm hadrons}) - \Gamma(\chi_{c1} \to {\rm hadrons})$ \cite{chiqcd}.
Here the strong coupling constant is set to be
$\alpha_s(m_c)=0.29 \pm 0.02$.}
A comparison of $\Gamma_{\gamma\gamma}(\chi_{c2})$
with the two-gluon width will provide a way to
study the validity of perturbative QCD corrections for quarkonium
decays.

\indent
   Measurements of two-photon decay widths for
charmonium states are difficult
because of their small production cross section
and small detection efficiencies. To date, several
experiments have reported
the observation of two-photon production of the $\chi_{c2}$ 
in the decay channel $\chi_{c2} \rightarrow \gamma
J/\psi$, $J/\psi \rightarrow \ell^{+}\ell^{-}$, where $\ell = e$ or $\mu$
 \cite{exps}.
This channel is suitable for the experimental determination of
$\Gamma_{\gamma\gamma}(\chi_{c2})$, since the decay branching
fraction is known with a relatively small error,
${\cal B}(\chi_{c2} \to \gamma J/\psi)
{\cal B}(J/\psi \to \ell^+\ell^-) = (1.59 \pm 0.13)\%$
for both lepton families \cite{pdb}.
However, the two-photon decay width results obtained from
previous measurements with this process seem to be systematically
larger than those from $p\bar{p} \to \chi_{c2} \to \gamma \gamma$
experiments \cite{pbarp}. Further studies with high statistics
data samples are needed to clarify the situation.

\indent
Recently, the CLEO collaboration has reported a measurement
of $\Gamma_{\gamma\gamma}(\chi_{c2})$ in
the $\pi^+\pi^-\pi^+\pi^-$ final
state \cite{cleo4pi}. Although the branching fraction
of this decay mode,
${\cal B}(\chi_{c2} \to \pi^+\pi^-\pi^+\pi^-)=(1.2 \pm 0.5)\%$  \cite{pdb}, is
comparable to that of the $\gamma J/\psi \to \ell^+\ell^-\gamma$ mode,
its large error precludes a 
precise determination of $\Gamma_{\gamma\gamma}(\chi_{c2})$.

\indent
We have measured $\chi_{c2}$ production in two-photon
processes using the decay channel $\chi_{c2} \rightarrow \gamma
J/\psi$, $J/\psi \rightarrow \ell^{+}\ell^{-}$.  The results are based
on a 32.6~fb$^{-1}$ data sample collected with the Belle detector.

\section{Experimental data and detector system}
The experiment was performed with the Belle detector \cite{Belle}
at the asymmetric $e^{+}e^{-}$ collider KEKB, 
where an 8.0~GeV $e^-$ beam collides with a 3.5~GeV $e^+$ beam
with a crossing angle of 22~mrad. We use a 29.6~fb$^{-1}$ 
sample of data collected
at the c.m. energy corresponding to the peak of the $\Upsilon(4S)$
resonance (10.58~GeV) and a 3.0~fb$^{-1}$ sample collected
60~MeV below the peak.

\indent
The basic topology of events that
we select is two tracks of opposite
charge and a photon. The recoiling $e^{+}$ and $e^{-}$ 
are not tagged in order to select
quasi-real two-photon collisions with high
efficiency.  Events induced by highly virtual photons
(i.e., photons with high-$Q^2$) are effectively rejected by a
strict transverse momentum($p_t$) requirement applied to the 
$\chi_{c2}$ daughter particles, as described in the
following section.\footnote{$Q^2$ is defined as the negative of the
invariant mass squared of a virtual incident photon.
It is approximately equal to $|p_t|^2$ of the virtual photon
with respect to the $e^+e^-$ beam axis.}

\indent
 The charged track momenta are measured with a cylindrical
drift chamber (CDC) located in a uniform 1.5~T magnetic field.
Track trajectory coordinates near the
collision point are provided by a
silicon vertex detector (SVD).  Photon detection and
energy measurements are performed with a CsI electromagnetic
calorimeter (ECL).
The resolutions of track momentum and photon energy 
measurements are 0.4\% for the leptons,
which, for the signal process, have typical
$p_t$ values of 1.4~GeV/$c$, and 2.0\% for the photons,
with typical energies of 0.4~GeV.  The magnet
return iron is instrumented to form
the $K_L$ and muon detector (KLM), which detects muon tracks 
and provides trigger signals.

\indent
The majority of the signal events are triggered
by two-track triggers that require at least
two tracks with transverse
momenta larger than 0.2 GeV/$c$ detected in the CDC
in  coincidence with matching signals
from TOF counters and trigger scintillation counters,
isolated cluster or energy-sum signals from the ECL, 
or muon tracks in the KLM.
A constraint on the opening angle in the plane transverse 
to the $e^+$ beam axis ($r\varphi$ plane), $\Delta \varphi >
135^{\circ}$, is applied at the trigger level.  
The signal inefficiency due to this trigger
constraint is negligibly small because the
final-state leptons tend to be back-to-back in the
$r\varphi$ plane due to the kinematic
properties of our selected final states,
namely the strict $p_t$ requirement on the $\chi_{c2}$ and
the small mass difference between the $\chi_{c2}$ and $J/\psi$.
The events from the $J/\psi \to e^+e^-$ mode are
efficiently triggered by a total energy trigger
derived from the ECL with a threshold set at 1.0~GeV.

\indent
 The main backgrounds are leptonic final states
from QED processes such as $e^+e^- \to e^+e^-\ell^+\ell^- \gamma$.
The $E/p$ information, which is the ratio of the energy
deposit in the ECL to the track's momentum, is used 
to identify  the leptons and eliminate small backgrounds that
contain hadron tracks.

\section{Event selection}
The event selection criteria are as follows: (1)
Exactly two oppositely charged tracks reconstructed by the CDC,
where both tracks satisfy the following laboratory
frame conditions:
$-0.47 \leq \cos \theta  \leq +0.82$, where $\theta$
is the polar angle;
$ p_t \geq 0.4~{\rm GeV/}c$;
$|dr| \le1$~cm,
$|dz| \le 3$~cm, 
where $(|dr|,\ dz)$
are the cylindrical coordinates of the track's point
of closest approach to the nominal collision point in the
$r\varphi$ plane; 
$|\Delta dz| \le 1$~cm, where
$\Delta dz$ is the difference between the $dz$'s of the two
tracks;
and  no other well reconstructed tracks 
with $p_t$ higher than
0.1~GeV/$c$.
(2) The opening angle ($\alpha$) of the two tracks
satisfies $\cos \alpha >-0.997$.
(3) There is just one  electromagnetic cluster 
in the ECL with an energy $E_{\gamma} \geq
0.2$~GeV and isolated from the nearest charged
track by an angle greater than $18^{\circ}$.
(4) The scalar sum of the momenta of the two charged tracks
is less than 6~GeV/$c$, and the invariant mass of the two tracks
is between 1.5 and 4.5~GeV/$c^2$.
(5) The total energy deposited in the ECL is less than 6~GeV.
(6) The absolute value of the total transverse momentum vector
in the c.m. frame of the $e^+e^-$ beams,
$|{\bf p}_t^{*\rm tot}|=|{\bf p}_t^{*+}+{\bf p}_t^{*-}+{\bf
p}_t^{*\gamma}|$, is less than 0.15~GeV/$c$, while
that for the two tracks only, $|{\bf p}_t^{*+}+{\bf p}_t^{*-}|$,
is larger than 0.10~GeV/$c$,
where ${\bf p}_t^{*+}$, ${\bf p}_t^{*-}$ and ${\bf p}_t^{*\gamma}$
are measured transverse momentum vectors (defined as
two-dimensional momentum vectors projected onto the plane 
perpendicular to the beam axis in the $e^+e^-$ c.m. system)
for the positive track,
the negative track and the photon, respectively.
(7) For electron pairs, both tracks are required to have
$E/p \geq 0.8$; for muon pairs, both tracks are required to have
$E/p \leq 0.4$.

\indent
Selection criterion (2) rejects cosmic-ray
backgrounds.  Criterion (6)  rejects
two-photonic lepton-pair production events with
radiation from a recoil electron or with a fake photon,
which tend to populate the region $|{\bf p}_t^{*+}+{\bf
p}_t^{*-}| \approx 0$.
The scatter plot in the
$|{\bf p}_t^{*+}+{\bf p}_t^{*-}|$-$|{\bf p}_t^{*\rm tot}|$ plane
before the application of requirement (6) is shown in Fig.~1(a),
where two separate clusters of events at
$|{\bf p}_t^{*+}+{\bf p}_t^{*-}| \approx 0$ and
$|{\bf p}_t^{*\rm tot}| \approx 0$ are apparent. 
The latter cluster corresponds
to exclusive $\ell^+\ell^-\gamma$ final states
produced by two-photon collisions.
The distribution from the signal events obtained from
the Monte Carlo (MC) simulation (described in Sect.~5) is shown
in Fig.~1(b).

\indent
We correct the absolute momenta of detected electrons
or positrons~ for bremsstrahlung in $e^+e^-\gamma$ event
candidates.  If photons of energy
between 0.02 and 0.2~GeV are present within a cone of half-angle
$3^{\circ}$ around the electron direction,
the energy of the most energetic photon
in the cone is added to the absolute momentum of the
track. This correction is also effective for partially
compensating for the
radiative-decay events of $J/\psi$, $J/\psi \to  e^+e^-\gamma$.

\begin{figure}[t]\centerline{\epsfxsize=6cm \epsfbox{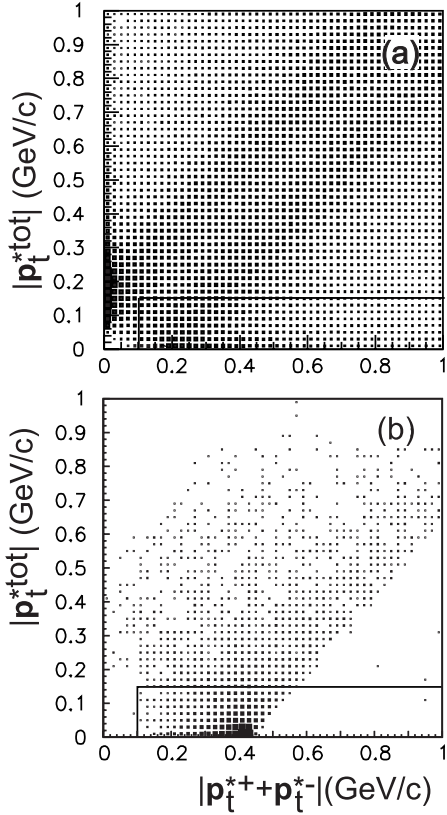}}
\def\thefigure{1}\caption{Scatter plots for the absolute values
of the two kinds of vector sums of the transverse momenta:
the horizontal axis for the sum of the two charged particles,
and the vertical axis of all three particles including
the photon: (a) for real data and (b) for Monte Carlo
events of the signal process. Straight lines show the
selection requirements.}\label{f1}\end{figure}

\section{Derivation of the number of signal events}
Figure~2  shows a scatter plot of the invariant mass of the
two tracks ($M_{+-}$) versus the invariant-mass difference
$\Delta M = M_{+-\gamma}-M_{+-}$ for the selected events,
where $M_{+-\gamma}$ is the invariant
mass of all three particles. A clear concentration is observed
around  $M_{+-}=3.097$~GeV/$c^2$
and $\Delta M=0.459$~GeV/$c^2$,
the signal region for $\chi_{c2} \to \gamma J/\psi$.

\indent
The mass difference
distribution is shown in Fig.~3(a) for the events falling within
the $J/\psi$ signal mass region
$3.06 \leq M_{+-} \leq 3.13$~GeV/$c^2$.
After the final selection requirement for signal candidate events,
$0.42 \leq \Delta M \leq 0.49$~GeV/$c^2$,
is applied, 176 events remain.  Of these, 82 events have
electron pairs and 94 have muon pairs. 
The contribution of $\chi_{c1}$ production, which
would peak at $\Delta M=0.42$~GeV/$c^2$, is estimated to be
less than one event, as expected from the suppression of
spin-1 meson production in
quasi-real two-photon collisions \cite{betsu}.

\begin{figure}[t]\centerline{\epsfxsize=9cm \epsfbox{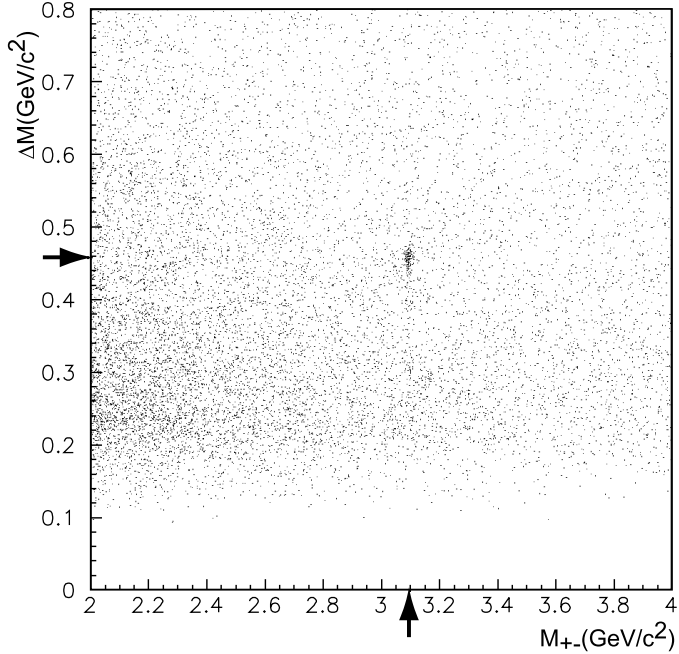}}
\def\thefigure{2}\caption{Scatter plot of the 
invariant mass $M_{+-}$ of the two-track system versus the mass
difference $\Delta M = M_{+-\gamma}-M_{+-}$. 
The arrows indicate the nominal
$J/\psi$ mass and the $\Delta M$ value
expected for $\chi_{c2}\to\gamma J/\psi$.}\label{f2}\end{figure}

\begin{figure}[t]\centerline{\epsfxsize=10cm \epsfbox{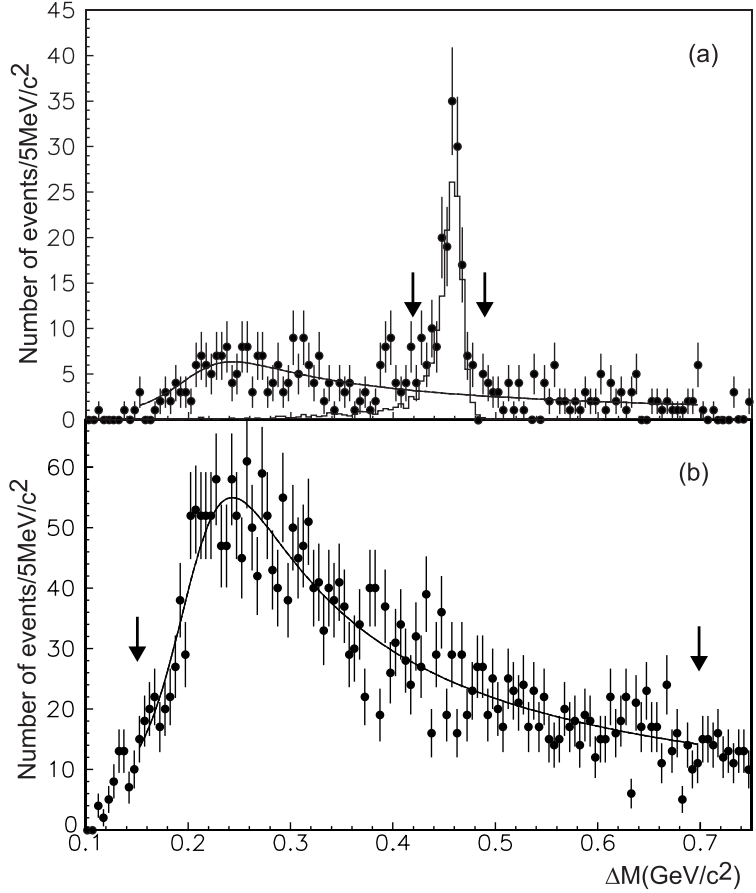}}
\def\thefigure{3}\caption{The mass difference distributions
for (a) events in the $J/\psi$-mass region (closed circles with
error bars) and (b) sideband events.  The curves
in (a) and (b) indicate the results of
the fits that are used to determine the
background contribution in the signal region.
The histogram in (a) shows the distribution of the
signal MC events normalized to the observed
signal.  The arrows show the signal region (a) and
the background control region (b).
}\label{f3}\end{figure}

\indent
 The $\Delta M$ distribution of events in the $J/\psi$-mass
sideband regions
($2.65<M_{+-}<3.00$~GeV/$c^2$ and $3.15<M_{+-}<3.50$~GeV/$c^2$)
is used to determine the background contribution
(Fig.~3(b)).
The $\Delta M$ distribution for $J/\psi$ sideband events
between $\Delta M = 0.15$ and 0.70~GeV/$c^2$ is fitted by
an appropriate function\footnote{ We used the following
empirical function for the fit of the background shape:\\
$A(\Delta M-a)^{-b}/(1+e^{-c(\Delta M-d)})$, where $a$, $b$, $c$ and
$d$ are free parameters, and $A$ the normalization parameter.
We have confirmed that the change of the $\Delta M$ distribution shape
is small at different $M_{+-}$ points in the sideband region.}
to obtain the shape of the background distribution.
We normalize this to the $\Delta M$
distribution of events in the $J/\psi$ signal region, and
determine
the number of background events in the $\chi_{c2}$ signal region
to be $40.0 \pm 2.7$ events.
This normalization fit uses only
data in the $\chi_{c2}$ sideband regions,
$0.15<\Delta M<0.29$~GeV/$c^2$ and $0.50<
\Delta M<0.70$~GeV/$c^2$.  Since we expect $9\pm7$
events around $\Delta M=0.32$~GeV/$c^2$
from the $\chi_{c0}$ \cite{pdb,cleo4pi}, we
avoid that mass region.  This expected $\chi_{c0}$ yield
is consistent with the small excess of events
seen near $\Delta M = 0.32$~GeV/$c^2$ in Fig.~3(a).
We determine the number of $\chi_{c2}$ signal events
to be $136.0 \pm 13.3$ after subtracting the number
of background events from the total
in the signal region.

We find that the
events in $\Delta M$ sideband regions are 
dominated by non-$J/\psi$ backgrounds since the
quantity and shape of the distribution agree with that of
the $J/\psi$ sideband.  The 368 events in the 
$0.42 \le \Delta M \le 0.49$~GeV/$c^2$ region
in Fig.~3(b) consist of 176 electron pairs
and 192 muon pairs.  These backgrounds are consistent with
higher-order QED events such as $e^+e^- \to e^+e^-\ell^+ \ell^-
\gamma$, which would give comparable numbers of events with
electron and muon pairs.  In contrast, hadron production (in
which pions would fake muons) would
give primarily events containing muon pairs.
Since a complete calculation of this process that takes
interference effects into account is not available, we cannot 
estimate the background yield theoretically.

\indent
Figure~4 shows distributions for $\Delta \varphi$,
$|{\bf p}_t^{\rm *tot}|$,
$|\cos \theta_{\gamma}^{+-\gamma}|$ and $\cos \theta_{-}^{+-}$
for the final candidate events,
where $\Delta \varphi$ is the azimuthal angle 
difference between the two lepton's momenta in the laboratory frame,
and $|{\bf p}_t^{\rm *tot}|$ is the transverse momentum of the
$\ell^+\ell^-\gamma$ system in the c.m. frame of the $e^+e^-$ beams. 
$\theta_{\gamma}^{+-\gamma}$ and $\theta_{-}^{+-}$ are the polar
angles of the photon in the $\ell^+\ell^-\gamma$ c.m. frame
and of the negatively charged lepton in the  $\ell^+\ell^-$
c.m. frame, respectively, where the polar angles are measured
with respect to the incident $e^-$ direction.
The experimental distributions are compared with the
sum of the signal MC events, described in the next section,
and the expected background contributions determined from
the $J/\psi$ sideband events, normalized to the observed numbers
of events.

\indent
It is apparent from Fig.~4(a) that the final-state leptons
have a back-to-back topology in the $r\varphi$ plane, and the
experimental data are consistent with the MC distribution.
The data in Fig.~4(b) show a peak at very small $p_t$
values ($|{\bf p}_t^{*{\rm tot}}| < 30$~MeV/$c$)
of the final-state system that is a typical feature of
exclusive production in two-photon collisions;
the data are in good agreement with the MC prediction.
The distribution for sideband events has a less prominent
concentration near $|{\bf p}_t^{*{\rm tot}}|=0$.
In the angular distributions of Figs.~4(c) and 4(d),
the data are consistent with the MC.

\begin{figure}[t]\centerline{\epsfxsize=9cm \epsfbox{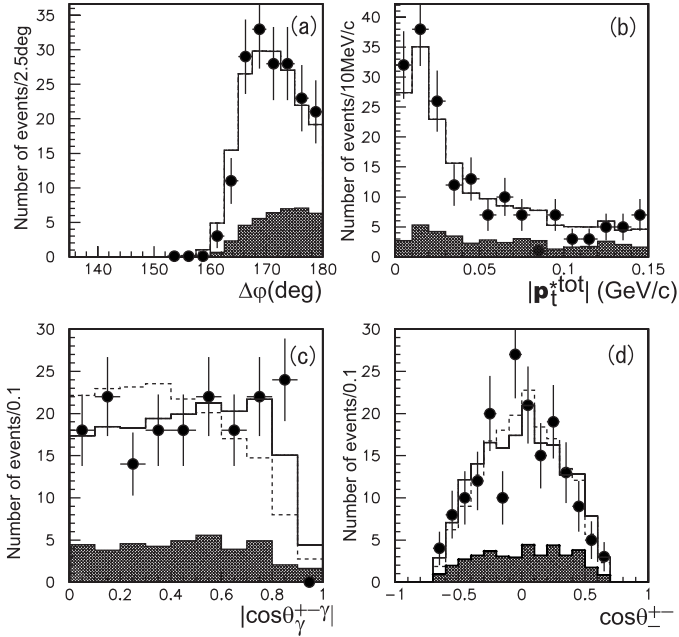}}
\def\thefigure{4}\caption{Comparison of the final samples 
(closed circles with error bars, backgrounds included)
with the sum of the signal MC events (open histogram)
and estimated background contributions (hatched region)
for: (a) $\Delta \varphi$;
(b) $|{\bf p}_t^{*{\rm tot}}|$;
(c) $|\cos \theta_{\gamma}^{+-\gamma}|$; and (d) $\cos
\theta_{-}^{+-}$.
There are no entries in the $\Delta\varphi < 135^\circ$ 
region in (a).
The dashed histograms in (c) and (d) show the distribution for 
the pure helicity=0 production case (see the text in Sect.~5).}
\label{f4}\end{figure}

%%%%%%%%%%%%%%%%%%%SLO

\section{Monte Carlo calculations}
We used Monte Carlo (MC) simulated
$e^+e^- \to e^+e^- \chi_{c2}$, $\chi_{c2} \to \gamma J/\psi$,
$J/\psi \to \ell^+\ell^-$ ($\ell= e$ or $\mu$) events to calculate the
efficiency for the signal process.
The TREPS MC program \cite{treps} is used for the
event generation.
The effects of $J/\psi$ radiative decays are
modeled with the PHOTOS \cite{photos} simulation
code, which generates photon radiation from a final-state lepton
generated by TREPS with a probability determined by a QED
calculation.  All of the final-state particles in the MC events
are processed by the full detector simulation program.

\indent
  We assume that the  $\chi_{c2}$ decay to the $\gamma J/\psi$
final state is an $E1$ transition, since  
experimental observations indicate that this transition
dominates the decay \cite{exE1}.
Since the helicity state of $\chi_{c2}$ produced in
two-photon collisions is not known, we assume a pure $\lambda =
2$ state  \cite{revpop}, where $\lambda$ is the helicity of
$\chi_{c2}$ with respect to the $\gamma\gamma$-incident axis.
The measurement of the polar-angle distribution
of the photon shown in Fig.~4(c) can be used to evaluate
the possible contribution from a $\lambda = 0$ component.
The $\lambda=2$ component produces
a $\cos \theta_{\gamma}^{+-\gamma}$ distribution
that is proportional to
$[1 + \cos^2 \theta_{\gamma}^{+-\gamma}]$, whereas the
$\lambda=0$ component is proportional to
$[5 -3\cos^2 \theta_{\gamma}^{+-\gamma}]$.
The dashed histograms in Figs.~4(c) and 4(d)
show the expected distributions from the pure $\lambda=0$ state.
The present experimental
$\cos \theta_{\gamma}^{+-\gamma}$ distribution has 
better agreement with a pure $\lambda = 2$ hypothesis
($\chi^2/dof=12.0/9$) than that for pure $\lambda=0$
($\chi^2/dof=45.0/9$), or any other
mixture of the two helicity states.  Here, each $\chi^2/dof$ is
calculated from the $|\cos \theta_{\gamma}^{+-\gamma}|$ distribution
divided into ten bins, where the total event number in the expected
signal distribution is fixed to the number of observed signal events.

\indent
The trigger efficiency
is experimentally determined using Bhabha and $\mu^+\mu^-$
events that are collected with two or more different redundant
triggers, as described above in Sect.~2.
We estimate the probability for the signal-process
events that survive the selection criteria to pass the
trigger conditions to be $(99 \pm 1)$\% ($(94 \pm 3)$\%) for
$ee\gamma$ ($\mu\mu\gamma$) events; and $(96 \pm 2)$\% in
average.  
This estimation agrees with the results of a trigger
simulation that is applied to the signal MC events. 
The trigger efficiency is also confirmed by the experimental 
yield of  two-photonic lepton-pair events, $e^+e^- \to
e^+e^-\ell^+\ell^-$, which agrees with the expectation from 
a QED calculation \cite{bdk}.

\indent
The ECL photon energy resolution was studied by comparing
experimental and MC mass-difference distributions for
$D^{*0} \to D^0 \gamma$ decays in $e^+e^-$ annihilation events.
We find
that the photon energy resolution is 1.3 times the
MC prediction.  The same tendency is confirmed in the
$\chi_{c1} \to \gamma J/\psi$ samples from $B$-meson decays and
$\eta \to \gamma\gamma$ samples from two-photon collisions.
If we use a correspondingly wider $\Delta M$ distribution for 
the $\chi_{c2} \to \gamma J/\psi$, it
decreases the efficiency of the
$\Delta M$ selection by 3.8\% from the MC-determined value.
We take this effect into account
as a correction for the efficiency and assign
a systematic error of the same size ($\pm 3.8\%$).  
We confirm that the observed position and width
of the signal peak in the data are consistent
with their expected values.

\indent
When the trigger effects and the difference between
the data and MC photon energy resolutions are taken into
account, the overall efficiency is found to be
6.6\%.  This is the average of the $e^+e^-$ and $\mu^+\mu^-$
decay channels;
the ratio of the efficiencies for the
two lepton species is $ee/\mu\mu = 0.70\pm0.03$.
The lower efficiency for $e^+e^-$ is due to 
the occasional presence of extra
high-energy photons from radiative $J/\psi$ decays
and electron bremsstrahlung.
When the $ee/\mu\mu$ ratio for the background component
determined from the $J/\psi$ sideband events is taken into
account, we expect the event yields
in the final samples to have a ratio $ee/\mu\mu = 0.74\pm0.04$.
This value is consistent with the observed ratio for the
final experimental samples: $0.87\pm0.13$.

\indent
The two-photon luminosity function is also calculated by  
the TREPS program \cite{treps}.  For consistency,
the same upper cutoff value of the photon $Q^2$,
$Q^2_{\rm max} = 1.0$~GeV$^2$, and the same vector-meson
pole effect are used in the calculation of
the luminosity function and in the event generation.
The uncertainty in the luminosity function
due to the vector-meson pole effect (we adopt the $J/\psi$ mass)
is small, about 2\%, since we apply a strict
$|{\bf p}_t^{\rm *tot}|$ requirement that is well below the
mass scale of the vector mesons.
The $Q^2_{\rm max}$ value does not affect
the product of the luminosity
function and the detection efficiency, since it 
is chosen to be large enough to cover the acceptance 
of our $|{\bf p}_t^{\rm *tot}|$ requirement.
The value of the TREPS luminosity function is
compared with that obtained from a full-diagram calculation of
the $e^+e^- \to e^+e^-\mu^+\mu^-$ process \cite{bdk} in the small
$|{\bf p}_t^{\rm *tot}|$ region.  
From the difference of the two results, the systematic
error for the luminosity function is estimated to be 5\%,
which includes ambiguities from the choice of the form factor
and the finite $|{\bf p}_t^{\rm *tot}|$ requirement.

\section{Results and discussion}
The two-photon decay width of the $\chi_{c2}$
is related to the signal event yield as
\begin{eqnarray}
\frac{\rm Yield}{\int {\cal L}dt} &=& 20\pi^2
\frac{L_{\gamma\gamma}(m_{\chi_{c2}})\eta}{(c/\hbar)^2 m_{\chi_{c2}}^2}
\Gamma_{\gamma\gamma}(\chi_{c2}) {\cal B}(\chi_{c2}
\to
\gamma J/\psi){\cal B}(J/\psi \to \ell^+\ell^-) \nonumber \\
&=& (0.309~{\rm fb/eV}) \times 
\Gamma_{\gamma\gamma}(\chi_{c2}) {\cal B}(\chi_{c2}
\to
\gamma J/\psi){\cal B}(J/\psi \to \ell^+\ell^-), \nonumber
\end{eqnarray}
where 
$\int {\cal L}dt$ is the integrated luminosity,
$\eta$ is
the efficiency, 
$m_{\chi_{c2}}$(=3.556~GeV/$c^2$) is
the $\chi_{c2}$ mass and 
$L_{\gamma\gamma}(m_{\chi_{c2}})(=7.75 \times 
10^{-4}$~GeV$^{-1}$) is the two-photon luminosity function
at the $\chi_{c2}$ mass.
The total width of $\chi_{c2}$ ($2.00 \pm 0.18$~MeV \cite{pdb})
is much smaller than the present
$\Delta M$ resolution ($\sim 9$~MeV), and does not affect
the present measurement.\\

\indent
The observed number of events, $136.0 \pm 13.3$,
implies the result
$\Gamma_{\gamma\gamma}(\chi_{c2}){\cal B}(\chi_{c2}\\ 
\to \gamma J/\psi)
{\cal B}(J/\psi \to \ell^+\ell^-) = 13.5 \pm 1.3 \pm 1.1$~eV,
where the first and second errors are statistical and
systematic, respectively. This result corresponds to
$\Gamma_{\gamma\gamma}(\chi_{c2}){\cal B}(\chi_{c2} \to \gamma J/\psi)
= 114 \pm 11(stat.) \pm 9(syst.) \pm 2(B.R.)$~eV or 
$\Gamma_{\gamma\gamma}(\chi_{c2}) = 0.85 \pm 0.08(stat.) \pm
0.07(syst.) \pm 0.07(B.R.)$~keV, where the last errors correspond to 
the uncertainties of the branching ratios, 
${\cal B}(\chi_{c2} \to \gamma J/\psi) = (13.5 \pm 1.1)\%$ and 
${\cal B}(J/\psi \to \ell^+\ell^-)=(11.81 \pm 0.20)\%$ \cite{pdb}.
The systematic error has contributions from
the trigger efficiency (2\%), lepton
identification efficiency (1.5\%), photon detection
efficiency (2\%),
inefficiency due to fake photons (less than 2\%),
$J/\psi$ detection efficiency (2\%),
the $\Delta M$ cut efficiency (3.8\%), background
subtraction (2.3\%),
the luminosity function (5\%) and other sources (less than 3\%);
these total 8\% when combined in quadrature.

\indent
The error in the background subtraction is derived from
the difference in signal yields in the
$\Delta M$ signal region between the present method
(counting the events in the signal region and subtracting the
background contribution)
and an alternative method in which the signal and the background
components
are simultaneously fitted to the $\Delta M$ distribution with
all the shape and size parameters for the background and signal
distributions allowed to float, with 
the Crystal-Ball line shape \cite{cbline}
used for the signal distribution.
The error of the background normalization is also combined with
this error.

\indent
The inefficiency
due to an extra (fake) photon with $E>0.2$~GeV is estimated to be
less than 2\% from an
experimental study  of the $p_t$-balanced $e^+e^- \to
e^+e^-\mu^+\mu^-$ process.

\indent
The uncertainty due to the assumption that the
$\chi_{c2}$'s are produced in a pure $\lambda=2$ state and  
decay via
pure $E1$ transitions is not included in the
systematic error. 
Production in the $\lambda=0$ state
at the 10\% level would increase
the detection efficiency by 7\% and decrease the measured
$\Gamma_{\gamma\gamma}(\chi_{c2})$ value by the same amount.
Meanwhile, a small mixture of the $M2$ transitions as
has been indicated by a measurement,
$M2/E1 \simeq -0.093^{+0.039}_{-0.041}(=a_2(\chi_{c2}))$
in amplitude \cite{exE1}, gives
only a 2\% effect on the efficiency.

\begin{figure}[t]\centerline{\epsfxsize=9cm \epsfbox{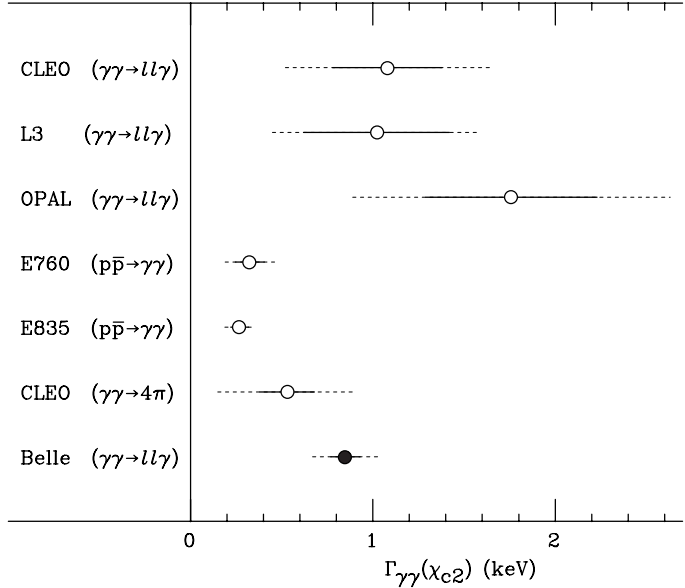}}
\def\thefigure{5}\caption{Comparison of the Belle
result for $\Gamma_{\gamma\gamma}(\chi_{c2})$
value with those from previous measurements[5$-$7].
The solid error bars show the statistical errors. The length
of the dashed part in each error bar corresponds to the
size of the systematic error, including the branching ratio
uncertainty.}
\label{f5}\end{figure}

\indent
The Belle result for $\Gamma_{\gamma\gamma}(\chi_{c2})$ 
is compared with those from
previous experiments \cite{exps,pbarp,cleo4pi} in Fig.~5.
The present result has the smallest statistical
and systematic errors of all the two-photon measurements
and is consistent with the previous
two-photon results. However, it is larger 
than the $p\bar{p}$ results.
A review of the experimental results of various
branching ratios of $\psi(2S)$ and $\chi_c$ decays \cite{branch}
suggests that this discrepancy may come from
incorrect values of
${\cal B}(\chi_{c2} \to \gamma J/\psi)$ and
${\cal B}(\chi_{c2} \to p \bar{p})$ that are 
used for the derivation of
$\Gamma_{\gamma\gamma}(\chi_{c2})$ in these experiments.

\section{Conclusion}
We have measured $\chi_{c2}$ production from two-photon collisions
with a 32.6~fb$^{-1}$ data sample collected with the Belle
detector at the KEKB $e^+e^-$ collider, using the decay mode
$\chi_{c2} \to \gamma J/\psi$, $J/\psi \to \ell^+\ell^-$.
We find $136.0 \pm 13.3$ signal
events after background subtraction.
The observed polar-angle distributions of the photon and leptons are
consistent with those expected from the production of
$\chi_{c2}$ in the pure helicity~2 state.
The product of the two-photon decay width of $\chi_{c2}$
and branching fractions,
$\Gamma_{\gamma\gamma}(\chi_{c2}){\cal B}(\chi_{c2} \to \gamma J/\psi)
{\cal B}(J/\psi \to \ell^+\ell^-)  = 13.5 \pm 1.3(stat.)
\pm 1.1(syst.)$~eV, is obtained.  This result corresponds to
$\Gamma_{\gamma\gamma}(\chi_{c2})
= 0.85 \pm 0.08(stat.) \pm 0.07(syst.) \pm 0.07(B.R.)$~keV, 
where the product of the branching fractions, ${\cal
B}(\chi_{c2} \to \gamma J/\psi)
{\cal B}(J/\psi \to \ell^+\ell^-)=(1.59 \pm 0.13)\%$ \cite{pdb}
is used.

\ \\
{\bf Acknowledgements}
\ \\
% Please paste this acknowledgement into your latex file.  
%***** Acknowledgments *****
We wish to thank the KEKB accelerator group for the excellent
operation of the KEKB accelerator.
We acknowledge support from the Ministry of Education,
Culture, Sports, Science, and Technology of Japan
and the Japan Society for the Promotion of Science;
the Australian Research Council
and the Australian Department of Industry, Science and Resources;
the National Science Foundation of China under contract No.~10175071;
the Department of Science and Technology of India;
the BK21 program of the Ministry of Education of Korea
and the CHEP SRC program of the Korea Science and Engineering Foundation;
the Polish State Committee for Scientific Research
under contract No.~2P03B 17017;
the Ministry of Science and Technology of the Russian Federation;
the Ministry of Education, Science and Sport of the Republic of Slovenia;
the National Science Council and the Ministry of Education of Taiwan;
and the U.S.\ Department of Energy.

\end{document}